\documentclass[aps,preprint,floatfix,nofootinbib,showpacs]{revtex4-1}
\pdfoutput=1
\usepackage{graphicx,color,rotating}
\usepackage{hyperref}
\usepackage{epsfig,color}

\begin{document}

\def\thefootnote{\fnsymbol{footnote}}

\title
{Resurgence of $Z'$ from the single electron-muon event at ATLAS}

\author{ Kingman Cheung$^{1,2,3}$, Wai-Yee Keung$^{4,1}$, and Po-Yan Tseng$^{2}$}
\affiliation{
$^1$ Physics Division, National Center for Theoretical Sciences, Hsinchu,
Taiwan \\
$^2$Department of Physics, National Tsing Hua University, 
Hsinchu 300, Taiwan \\
$^3$Division of Quantum Phases \& Devices, School of Physics, 
Konkuk University, Seoul 143-701, Republic of Korea \\
$^4$Department of Physics, University of Illinois at Chicago, IL 60607 USA
}

\date{\today}

\begin{abstract}
Inspired by the recent single $e^{\pm}\mu^{\mp}$
  event at 2.1 TeV invariant mass from the ATLAS at $\sqrt{s}=13$ TeV with 
  3.2 fb$^{-1}$ luminosity, we propose an explanation using a 
  $Z'$ gauge boson, which possesses lepton-flavor-changing 
  neutral currents originated from non-universal couplings to charged leptons.
  We assume that the left-handed charged-lepton 
  mixing matrix equals to the PMNS matrix and no mixing in the neutrino sector
  to make this phenomenological $Z'$ model more predictive.  
  There are
  indeed some parameter regions, where the $Z'$ can generate a large enough
  $e^{\pm}\mu^{\mp}$ production cross section, while at the same time 
  satisfies various observables from lepton-flavor violation and 
  other constraints from the LHC.
\end{abstract}

\maketitle

\section{Introduction}

The ATLAS collaboration recently reported the opposite-sign different-flavor
dilepton $e^{\pm}\mu^{\mp}$ pairs, using 3.2 fb$^{-1}$ data at
$\sqrt{s}=13$ TeV in Ref.~\cite{atlas_emu}. In the plot of the
spectrum of electron-muon invariant mass ($m_{e\mu}$), there is one
event at $m_{e\mu}=2.1$ TeV, where the expected background
is almost zero. The largest local
significance is $1.7\sigma$ at $m_{e\mu}=2.1$ TeV. From the difference 
between the observed and expected limits at $2.1$ TeV in Ref.~\cite{atlas_emu},
we estimate that the required cross section is around 1 -- 2 fb:
\begin{equation}
\sigma(pp \to X) \times  B(X \to e^{\pm}\mu^{\mp})\simeq 1-2\; {\rm fb}\;,
\end{equation}
through a heavy resonance with 2.1 TeV mass to explain the data.

In this work, we postulate that the excess is due to a gauge boson $Z'$
with a mass at $2.1$ TeV corresponding to an extra $U'(1)$. The particle
couples to both quarks and leptons, hence it can be produced by
quark-antiquark fusion at the LHC.  With regard to the leptonic
couplings, it violates the universality and has different strengths
for different flavors. The non-universal couplings to charged leptons
are also inspired from a recent observation of $b \to s l^+l^-$ in
Ref.~\cite{bsll}.

If we further assume that the mass matrix of the charged leptons is not
diagonal under the interaction basis and the couplings to $Z'$ are
non-universal, flavor-changing neutral currents (FCNC) can be induced
at tree level in the charged-lepton sector after diagonalizing their mass
matrices~\cite{ckm_zp,pmns_zp1,pmns_zp2,langacker}.
After a unitary transformation on the basis, non-zero $Z'e\mu$ coupling can
be generated. However, complete informations of the unitary
transformation on left- and right-handed charged leptons, $U_{lL}$ and
$U_{lR}$, are still ambiguous, since the neutrino oscillation
observations always measure the product of the left-handed charged
lepton and neutrino unitary matrices, i.e the PMNS matrix is
$U_{PMNS}=U^{\dagger}_{lL}U_{\nu}$.

In order to make the couplings of $Z'$ more predictive, we further
postulate that the PMNS lepton-mixing matrix entirely comes from the
charged-lepton sector \cite{pmns_zp1,pmns_zp2}, i.e. $U_{\nu}={\bf 1}$
and $U^{\dagger}_{lL}=U_{PMNS}$.  Based on this framework, we stress
that the $Z'$ boson can generate large enough $\sigma(pp \to X) \times B(X
\to e^{\pm}\mu^{\mp})$, meanwhile still evades the constraints from
various kinds of observations.

We organize this work as follows. In Sec.~\ref{sii}, we 
introduce the notation, then consider possible constraints from other
leptonic and dijet channels at the LHC in Sec.~\ref{siii}. Bounds from
various low-energy observables relevant to the lepton sector itself, or
both lepton and quark sector will be considered in Sec.~\ref{siv}
and~\ref{sv}, respectively.  Numerical results are in
Sec.~\ref{svii}, and we summarize in Sec.~\ref{sviii}.

\section{Formalism}
\label{sii}

Here we trail the heavy resonance as a new gauge boson
$Z'$ from an extra $U'(1)$ additional to the SM gauge groups.  The
gauge couplings of the $Z'$ to different generations of fermions may
not be universal from some hints of the recent observations of $b \to
s l^+l^-$ in Ref.~\cite{bsll}. Flavor-changing neutral currents (FCNC)
can be induced at tree level in both quark and leptonic sectors after
diagonalizing their mass matrices \cite{ckm_zp,pmns_zp1,pmns_zp2}. We
follow the formalism in Ref.~\cite{ckm_zp}.  In the interaction
basis, the neutral-current Lagrangian from $Z'$ can be written as
\begin{equation}
\mathcal{L}_{\rm NC} = -g'J^{(2)\mu}Z'_{\mu}\;,
\end{equation}
where there is no mixing between $Z'$ and $Z$ boson from 
$SU(2)\times U(1)$ for simplicity, and $g'$ is the gauge coupling of 
$U'(1)$. The current associated with the $U'(1)$ is
\begin{equation}
\label{eq2}
J^{(2)}_{\mu}=\sum_{i,j} \bar{\psi}_i \gamma_{\mu}
\left[ {\epsilon^{\psi}_{L}}_{ij}P_L+{\epsilon^{\psi}_{R}}_{ij}P_R \right]\psi_j \;,
\end{equation}
where ${\epsilon^{\psi}_{{L,R}}}_{ij}$ 
are the chiral charges of $U'(1)$
with fermions $i$ and $j$ running over all quarks and leptons in the
interaction basis.

The $U'(1)$ charge assignment for left- and right-handed quarks are
universal, i.e ${\epsilon^{u}_{L,R}}=Q'^{(u)}_{L,R}\, diag(1,1,1)$,
${\epsilon^{d}_{L,R}}=Q'^{(d)}_{L,R}\, diag(1,1,1)$.  However, the
$U'(1)$ charges for the charged-lepton sector could be non-universal,
i.e. ${\epsilon^{l}_{L}}=diag(Q'^{(e)}_L, Q'^{(\mu)}_L,
Q'^{(\tau)}_L)$. Finally, for the $U'(1)$ charges for the right-handed
leptons, we simply assume they are zero, ${\epsilon^{l}_{R}}=0$.

The fermions in Eq.~(\ref{eq2}) in the interaction basis will be
rotated to the mass eigen-basis through a set of unitary matrices, e.g
${V_{u,d}}_{L}$, ${V_{u,d}}_{R}$ for left-, right-handed up- and
down-type quarks, respectively; $U_{lL}$, $U_{lR}$, and $U_{\nu}$ for
leptons and neutrinos. Therefore, the interactions between $Z'$ and
fermions in mass eigen-basis become
\begin{eqnarray}
\mathcal{L}_{\rm NC} &=& -g'Z'_{\mu} (\bar{u},\bar{c},\bar{t})_M \gamma^{\mu} 
(V^{\dagger}_{uL}\epsilon^u_LV_{uL}P_L+V^{\dagger}_{uR}\epsilon^u_RV_{uR}P_R)
\left( u, c, t \right)^T_M  \nonumber \\ 
&& -g'Z'_{\mu} (\bar{d},\bar{s},\bar{b})_M \gamma^{\mu} 
(V^{\dagger}_{dL}\epsilon^d_LV_{dL}P_L+V^{\dagger}_{dR}\epsilon^d_RV_{dR}P_R)
\left( d, s, b \right)^T_M  \nonumber \\ 
&& -g'Z'_{\mu} (\bar{e},\bar{\mu},\bar{\tau})_M \gamma^{\mu} 
(U^{\dagger}_{lL}\epsilon^l_LU_{lL}P_L+U^{\dagger}_{lR}\epsilon^l_RU_{lR}P_R)
\left( e, \mu, \tau \right)^T_M \;, \\
\label{eq4}
\Rightarrow \mathcal{L}_{\rm NC} &=& -Z'_{\mu} (\bar{u},\bar{c},\bar{t})_M \gamma^{\mu} 
(g^u_L P_L+g^u_R P_R)
\left( u, c, t \right)^T_M  \nonumber \\ 
&& -Z'_{\mu} (\bar{d},\bar{s},\bar{b})_M \gamma^{\mu} 
(g^d_L P_L+g^d_R P_R)
\left( d, s, b \right)^T_M  \nonumber \\ 
&& -Z'_{\mu} (\bar{e},\bar{\mu},\bar{\tau})_M \gamma^{\mu} 
(g^l_LP_L+g^l_RP_R)
\left( e, \mu, \tau \right)^T_M \;,
\end{eqnarray}
where $g^{u,d,l}_{L,R}$ are $3\times3$ matrix describe the $Z'$
coupling to SM fermions. Since $\epsilon^u_{L,R}$,
$\epsilon^d_{L,R}$, and $\epsilon^l_R$ are
proportional to the identity matrix, no off-diagonal terms
will be generated after sandwiched by the unitary matrices. On the other
hand, since the diagonal elements in the $\epsilon^l_L$ are non-universal,
it will generate non-zero off-diagonal terms after sandwiched by 
the unitary matrices. 
The non-zero off-diagonal elements can 
induce the FCNC of  $Z'$.

In the leptonic sector, the PMNS matrix is
$U_{PMNS}=U^{\dagger}_{lL}U_{\nu}$ and we assume that all the neutrino mixings
come from the charged-lepton sector~\cite{pmns_zp1,pmns_zp2}, i.e
$U_{\nu}=${\bf 1}, then
\begin{equation}
V_{PMNS}=U^{\dagger}_{lL}\;.
\end{equation}
Therefore, the couplings of the 
left-handed leptons is $g^l_L=g'U_{PMNS}\epsilon^l_L U^{\dagger}_{PMNS}$, 
such that $g^l_L$ can be determined using the experimentally measured
$U_{PMNS}$ matrix and thus gives meaningful predictions.

\section{Constraints from $e^{\pm}\mu^{\mp}$, $e^+e^-$, $\mu^+\mu^-$, 
$\tau^+\tau^-$, and $jj$ production at the LHC}
\label{siii}

There are several constraints and upper limits for
$e^{\pm}\mu^{\mp}$~\cite{cms8TeVemu}, $e^{\pm}\tau^{\mp}$,
$\mu^{\pm}\tau^{\mp}$~\cite{atlas_emu}, $e^+e^-$,
$\mu^+\mu^-$~\cite{cms8TeVll,atlas8TeVll,atlas13TeVll}, and
$\tau^+\tau^-$ channels from ATLAS and CMS already. 
In Ref~\cite{atlas13TeVll}, 
the observed 95\% upper limits at
$\sqrt{s}=13$ TeV at $m_{Z'}=2.1$ TeV are $\sigma(pp\rightarrow
Z')\times B(Z'\rightarrow e^+e^-)\lesssim 1.5$ fb and
$\sigma(pp\rightarrow Z')\times B(Z'\rightarrow \mu^+\mu^-)\lesssim 2$
fb.  For channels of different flavors \cite{atlas_emu}, at 2.1 TeV,
$\sigma(pp\rightarrow Z')\times B(Z'\rightarrow
e^{\pm}\tau^{\mp})\lesssim 5$ fb, and $\sigma(pp\rightarrow Z')\times
B(Z'\rightarrow \mu^{\pm}\tau^{\mp})\lesssim 9$ fb.

The dijet limits from ATLAS~\cite{atlas13TeVdijet,atlas13TeVdijet2}
are about $\sigma(pp\rightarrow Z')\times B(Z'\rightarrow jj)\times
A\lesssim 0.5$ pb for a narrow-width $Z'$, and $\lesssim 1$ pb for
$\Gamma_{Z'}/m_{Z'}=0.15$ at $M_{Z'}\simeq 2.1$ TeV.  From the
CMS~\cite{cms13TeVdijet}, $\sigma(pp\rightarrow Z')\times
B(Z'\rightarrow jj)\times A\lesssim 1$ pb for the narrow-width case.
Here $A$ is the acceptance ratio due to selection cuts, and ranges
between $40-60\%$.

\section{constraints from the FCNC in the leptonic sector}
\label{siv}

In this section, we focus on the observables relevant to the
flavor-changing $Z'$-charged-lepton couplings, 
such as $\mu \to e\gamma$ or $\mu \to 3e$. 
The experimental limits from these processes are listed in
Table~\ref{tab:cx}.  

\subsection{$l_j \to l_i\gamma$}

The expression for the branching ratio $l_j \to l_i\gamma$ is~\cite{llgamma}
is given by
\begin{equation}
\label{eq7}
B(l_j \to l_i\gamma)=\frac{\alpha_e \tau_j m_j}{9(4\pi)^4}
\left(\frac{m_j}{m_{Z'}} \right)^4
\left( \left | \sum_k (g^l_L)_{jk}(g^l_L)_{ki}-\frac{3m_k}{m_j}
 (g^l_L)_{kj}(g^l_R)_{ki} \right |^2+(L\leftrightarrow R) \right) \;,
\end{equation}
where fine structure constant $\alpha_e\equiv e^2/4\pi=1/137.036$ at
very low energy~\cite{pdg2015}, $m_{i,j,k}$ are mass of charged
leptons ($m_{e,\mu,\tau}=$0.0005, 0.10567, 1.77682 GeV), and
$\tau_{j}$ is the life time of the charged lepton $j$ 
($\tau_{\mu}=3.34\times10^{18}$, and
$\tau_{\tau}=4.42\times10^{11}$ GeV$^{-1}$~\cite{pdg2015} ).
Here we adopt the recent results from MEG
Collaboration~\cite{meg}, i.e. $B(\mu \to e\gamma)<4.2\times
  10^{-13}$ at 90\% CL. 

From the expression in Eq.(\ref{eq7}), if there are only left-handed
couplings $g^l_L$, the mass insertion only occurs at external lepton
legs in the Feynman diagram. However, when both left- and
right-handed couplings are nonzero, the mass insertion can happen in
the internal fermion loop in the Feynman diagram and so it could be of 
different flavor from the external leptons. In the latter case, mass ratio
$m_k/m_j$ in Eq.(\ref{eq7}) may enhance the decay rate of $l_j \to
l_i\gamma$.
For instance, among the current experimental limits the most
stringent one is from $\mu \to e\gamma$. If both left- and
right-handed couplings are nonzero, the diagram with the mass insertion
in the  $\tau$ running in the loop will be enhanced by the factor
$m_{\tau}/m_{\mu}$. Therefore, in order to dodge the experimental
limit of $B(\mu \to e\gamma)$, we assume $g^l_R=0$.

\subsection{$l_j \to l_il_k\bar{l_l}$}

The expressions for the branching ratios 
$l_j \to l_il_k\bar{l_l}$ are given by~\cite{llgamma} 
\begin{eqnarray}
B(l_j \to l_il_k\bar{l_l})&=&\frac{\tau_j m_j}{1536\pi^3}
\left(\frac{m_j}{m_{Z'}} \right)^4  \nonumber \\
&\times & 
\left( \left|(g^l_L)_{ij}(g^l_L)_{kl}+(g^l_L)_{kj}(g^l_L)_{il} \right|^2
+\left |(g^l_L)_{ij}(g^l_R)_{kl}\right |^2+ 
  \left|(g^l_L)_{kj}(g^l_R)_{il} \right|^2 
 +(L\leftrightarrow R) \right) \;, \nonumber  \\
B(l_j \to l_il_i\bar{l_l})
&=&\frac{\tau_j m_j}{1536\pi^3}
\left(\frac{m_j}{m_{Z'}} \right)^4
\left( 2 \left|(g^l_L)_{ij}(g^l_L)_{il} \right|^2+ 
 \left|(g^l_L)_{ij}(g^l_R)_{il} \right|^2+(L\leftrightarrow R) \right) \;.
\end{eqnarray}
The observed limit of $\mu^- \to e^-e^-e^+$ is less than $1.0\times
10^{-12}$~\cite{pdg2015}, which not only constrains the
flavor-changing coupling $(g^l_L)_{12}$, but also the
flavor-conserving one $(g^l_L)_{11}$.  So we have to suppress the
$Z'ee$ coupling as well in our numerical study.

\begin{table}[t!]
 \caption{\small \label{tab:cx}
Various experimental constraints coming from the LHC, rare lepton-flavor
violating decays, and $\mu$-$e$ conversions, as well as the predictions
of the benchmark point ($Z'$ M-1): (NH) 
$g'=1$, ${\epsilon^{u}_{L}}=-{\epsilon^{u}_{R}}= diag(0.2,0.2,0.2)$, 
${\epsilon^{d}_{L}}=-{\epsilon^{d}_{R}}=diag(0.2,0.2,0.2)$, 
${\epsilon^{l}_{L}}=1/10\times diag(-0.404,0.912,-0.064)$, 
${\epsilon^{l}_{R}}=0$. 
The total width of the $Z'$ is $\Gamma_{Z'}=40.7$ GeV, and the $Z'$ 
production cross section $\sigma(pp\to Z')=367$ fb at the 13 TeV LHC.}
\smallskip
\begin{ruledtabular}
\begin{tabular}{l|l|l}
 observable   & exp. & $Z'$ M-1  \\
\hline
$\sigma(pp\to Z') \times B(Z'\to e^{\pm}\mu^{\mp})$ [fb] &  1 $\sim$ 2~\cite{atlas_emu}     & 1.03 \\
$\sigma(pp\to Z') \times B(Z'\to e^+e^-)$ [fb]           &  $\lesssim$1.5~\cite{atlas13TeVll}  & $1.4\times 10^{-7}$  \\
$\sigma(pp\to Z') \times B(Z'\to \mu^+\mu^-)$ [fb]       &  $\lesssim$2~\cite{atlas13TeVll}    & 0.210 \\
$\sigma(pp\to Z') \times B(Z'\to \tau^+\tau^-)$ [fb]     &  -              & 0.060 \\
$\sigma(pp\to Z') \times B(Z'\to e^{\pm}\tau^{\mp})$ [fb] &  $\lesssim$5~\cite{atlas_emu}   & 0.782 \\
$\sigma(pp\to Z') \times B(Z'\to \mu^{\pm}\tau^{\mp})$ [fb] &  $\lesssim$9~\cite{atlas_emu} & 0.428 \\
$\sigma(pp\to Z') \times B(Z'\to jj)$ [fb]   &  $\lesssim$500~\cite{atlas13TeVdijet}              & 362 \\
\hline
$B(\mu \to e \gamma)$           & $<4.2\times 10^{-13}$~\cite{meg} & $4.4\times 10^{-13}$ \\
$B(\mu^- \to e^-e^-e^+)$        & $<1.0\times 10^{-12}$~\cite{pdg2015}     & $1.1\times 10^{-16}$ \\
$B(\tau \to \mu \gamma)$        & $<4.4\times 10^{-8}$~\cite{pdg2015}     & $1.2\times 10^{-13}$ \\
$B(\tau^- \to \mu^-\mu^-\mu^+)$ & $<2.1\times 10^{-8}$~\cite{pdg2015}     & $1.2\times 10^{-11}$ \\
$B(\tau^- \to \mu^-e^-e^+)$     & $<1.8\times 10^{-8}$~\cite{pdg2015}     & $2.7\times 10^{-11}$ \\
$B(\tau \to e \gamma)$          & $<3.3\times 10^{-8}$~\cite{pdg2015}     & $4.8\times 10^{-14}$ \\
$B(\tau^- \to e^-e^-e^+)$       & $<2.7\times 10^{-8}$~\cite{pdg2015}     & $1.5\times 10^{-17}$ \\
$B(\tau^- \to e^-\mu^-\mu^+)$   & $<2.7\times 10^{-8}$~\cite{pdg2015}     & $5.0\times 10^{-11}$ \\
\hline
$B(\mu{\rm Ti} \to e{\rm Ti})$    & $<6.1\times 10^{-13}$~\cite{muTieTi} & $0$  \\
$B(\mu{\rm Au} \to e{\rm Au})$    & $<7.0\times 10^{-13}$~\cite{pdg2015}      & $0$ \\
$B(\mu{\rm Al} \to e{\rm Al})$    & -                     & $0$  \\
\end{tabular}
\end{ruledtabular}
\end{table}

\section{constraints from the FCNC in the lepton-quark sector}
\label{sv}

In this section, we focus on the $\mu-e$ conversion processes in heavy
nuclei, which are relevant to the $Z'$-charged-lepton and $Z'$-quark
couplings. For the vector-like $Z'$ interactions, these processes will
be enhanced through coherent scattering with the entire nucleus,
therefore putting strong bounds on the $Z'$ couplings. The experimental limits
from these processes are listed in Table~\ref{tab:cx}.

\subsection{$\mu-e$ conversion: $\mu+N \to e+N$}
\label{mueconversion}

For coherent $\mu^- - e^-$ conversion only scalar and vector couplings 
contribute. In our $Z'$ model, there is only the vector contribution,
and no scalar couplings will be generated if RG running 
is restricted to QCD dressing only. 
The relevant expressions can be found in Ref.~\cite{muNeN1}
\begin{equation}
 B(\mu^-N \to e^-N)=\frac{p_eE_eG^2_F}{8\pi}
\left( |X_L(p_e)|^2+|X_R(p_e)|^2 \right)\frac{1}{\Gamma_{capt}}\, 
\end{equation}
where $p_e$ and $E_e$ is the momentum and energy of the electron, respectively,
$\Gamma_{capt}$ is the muon capture rate from the experiment, and
\begin{eqnarray}
X_L(p_e)&=& \left ( g^{(0)}_{LV}+g^{(1)}_{LV} \right)\,  ZM_p(p_e)
 + \left ( g^{(0)}_{LV}-g^{(1)}_{LV} \right )\, (A-Z)M_n(p_e) \;, \nonumber \\
X_R(p_e)&=&\left( g^{(0)}_{RV}+g^{(1)}_{RV} \right)\, ZM_p(p_e)+
 \left( g^{(0)}_{RV}-g^{(1)}_{RV} \right)\, (A-Z)M_n(p_e) \;, \nonumber
\end{eqnarray}
where $Z$ and $A$ are, respectively, the proton and nucleon numbers 
of the nucleus. The $M_{p,n}$ are the transition nuclear matrix elements.
Also, 
\begin{eqnarray}
g^{(0)}_{LV}&=&\frac{1}{2}\sum_{q=u,d,s}
\left( g_{LV(q)}G^{(q,p)}_V+g_{LV(q)}G^{(q,n)}_V \right) \;, \nonumber \\
g^{(0)}_{RV}&=&\frac{1}{2}\sum_{q=u,d,s}
\left( g_{RV(q)}G^{(q,p)}_V+g_{RV(q)}G^{(q,n)}_V \right) \;, \nonumber \\
g^{(1)}_{LV}&=&\frac{1}{2}\sum_{q=u,d,s}
\left( g_{LV(q)}G^{(q,p)}_V-g_{LV(q)}G^{(q,n)}_V \right) \;, \nonumber \\
g^{(1)}_{RV}&=&\frac{1}{2}\sum_{q=u,d,s}
\left( g_{RV(q)}G^{(q,p)}_V-g_{RV(q)}G^{(q,n)}_V \right) \;, \nonumber \\
\end{eqnarray}
for vector currents $G^{(u,p)}_V=G^{(u,n)}_V=2$,
$G^{(d,p)}_V=G^{(u,n)}_V=1$, and $G^{(s,p)}_V=G^{(s,n)}_V=0$ from
Ref.~\cite{muNeN2}. Comparing the effective operators in
Ref.~\cite{muNeN1} with Eq.(\ref{eq4}), the coefficients of these
operators can be written in terms of $Z'$ couplings
\begin{eqnarray}
g_{LV(q)}&=&\frac{\sqrt{2}}{m^2_{Z'} G_F}
(g^l_L)_{12}\left[ (g^q_R)_{11}+(g^q_L)_{11} \right] /2 \;, \nonumber \\
g_{RV(q)}&=&\frac{\sqrt{2}}{m^2_{Z'} G_F}
(g^l_R)_{12}\left[ (g^q_R)_{11}+(g^q_L)_{11} \right] /2 \;, \nonumber \\
g_{LA(q)}&=&\frac{\sqrt{2}}{m^2_{Z'} G_F}
(g^l_L)_{12}\left[ (g^q_R)_{11}-(g^q_L)_{11} \right] /2 \;, \nonumber \\
g_{RA(q)}&=&\frac{\sqrt{2}}{m^2_{Z'} G_F}
(g^l_R)_{12}\left[ (g^q_R)_{11}-(g^q_L)_{11} \right] /2 \;, \nonumber \\
\end{eqnarray}
where $q=u,d$.
Here we shall consider those experiments with three different nuclei
$N=\;^{27}{\rm Al},\;^{48}{\rm Ti},\;^{197}{\rm Au}$. Useful values
for these experiments are listed in Table 1 of Ref.~\cite{muNeN2}.

Note that in our $Z'$ model only the vector couplings in the quark
sector significantly contribute to the $\mu-e$ conversion. 
We can easily evade the current
experimental limits by choosing $U'(1)$ charges as
$\epsilon^u_L=-\epsilon^u_R$ and $\epsilon^d_L=-\epsilon^d_R$, such
that the $Z'$ couplings to quarks are almost axial-vector-like.
However, even under this $U'(1)$ charge assignment, it is still
possible that if the $Z'$ has non-universal couplings in the quark sector,
after the unitary transformation and rotating into the quark mass
basis, vector-like couplings can be induced. Then, the $Z'$ will
suffer from the strong limits of $\mu-e$ conversion.
Therefore, in order to escape from this once and for all, later in our
numerically analysis, the $Z'$ has universal couplings in quark sector, and
we assign opposite  $U'(1)$ charges to the left- and right-handed
quarks in order to evade the stringent $\mu-e$ conversion limits.

Another advantage of the assumption that the $Z'$ has universal couplings
in the quark sector is that we do not need to take into account the 
flavor-changing observables in the quark sector, such as 
$B-\bar{B}$ or $K-\bar{K}$ mixing.

Now we address the mechanism of the fermion mass
generation. For the quark sector in the scheme of the universal and
axial-vector-like couplings, the type-II model of two Higgs doublets
of opposite hypercharges and opposite $Z'$ charges is able to have
the gauge compatible Higgs-Yukawa couplings that generate the
required quark masses.  However, the lepton masses require more
technical structures in the Higgs sector due to the non-universal 
$U'(1)$ charges of leptons.  
In this paper, we are
only concerned about the phenomenological study of the $Z'$
interaction and put aside the Higgs interaction.

\section{numerical results}
\label{svii}

Here we shall demonstrate step by step how to assign the $U'(1)$ charges 
for the charge leptons and quarks, so as to make the model 
consistent with all the observables, and then to check if 
there is any parameter space left that can be tuned to generate large enough 
$\sigma(pp\to Z') \times B(Z'\to e^{\pm}\mu^{\mp})\simeq 1$ fb.

The PMNS matrix with the best-fit values of matrix elements is given in 
Particle Data Book~\cite{pdg2015} as 
(assuming zero values for the two Majorana CP violation phases):
\begin{eqnarray}
U_{PMNS}= \left(\begin{array}{ccc}
     0.822          &  0.548              & -0.0518+0.144 i        \\[2mm]
    -0.388+0.0791 i &  0.643+0.0528 i     &  0.653                 \\[2mm]
     0.399+0.0898 i & -0.528+0.0599 i     &  0.742   
             \end{array}\right)\, ,  \nonumber     \label{pmns_nh}
\end{eqnarray}
in the normal hierarchy(NH), and 
\begin{eqnarray}
U_{PMNS}= \left(\begin{array}{ccc}
     0.822          &  0.548              & -0.0525+0.146 i        \\[2mm]
    -0.380+0.0818 i &  0.634+0.0546 i     &  0.666                 \\[2mm]
     0.407+0.0895 i & -0.540+0.0597 i     &  0.729   
             \end{array}\right)\, ,  \nonumber     \label{pmns_ih}
\end{eqnarray}
in the inverse hierarchy(IH).

In Table~\ref{tab:cx}, we show an example of the $U(1)'$ charge
assignment for quarks and leptons, such that it can give a large enough cross
section for $\sigma(pp\to Z') \times B(Z'\to e^{\pm}\mu^{\mp})\simeq 1$
fb, and meanwhile satisfies the limits from all other observations.

The steps in assigning the $U(1)'$ charges are as follows.
\begin{itemize}
\item[(i)]
Considering in the leptonic sector very strong experimental limits come from
$B(\mu \to e \gamma)$ and $B(\mu^- \to e^-e^-e^+)$. The latter
limit can be satisfied by suppressing the $Z'ee$ coupling, i.e
$(g^l_L)_{11}$. The $Z'$ couplings are $g^l_L=g'U_{PMNS}\epsilon^l_L
U^{\dagger}_{PMNS}$, where
$\epsilon^l_L=diag(Q'^{(e)}_L,Q'^{(\mu)}_L,Q'^{(\tau)}_L)$. 
The coupling $(g_L^l)_{ij}$  depends linearly on
$ {Q'}_L^{(e)}, {Q'}_L^{(\mu)}, {Q'}_L^{(\tau)}$ with coefficients
$ ({\vec{A}}_{ij})_l$,
$$ 
(g_L^l)_{ij} =g' ({\vec{A}}_{ij})_l \ {Q'}_L^{(l)} \ ,                       
\hbox{ or } \ g'{\vec{A}}_{ij} \cdot {\vec{Q}}'_L   \ , 
$$
where
$$
  ({\vec{A}}_{ij})_l=  (U_{PMNS})_{il} (U^*_{PMNS})_{jl}  \ . 
  $$
We attempt to assign the $U(1)'$ charges, such that 
maximize the $e-\mu$ coupling $(g^l_L)_{12}$, while minimize the $e-e$ one
$(g^l_L)_{11}$ .
In NH, two $U(1)'$ charge assignments along the  directions
$\vec{A}_{12}\simeq(-0.319,0.353,-0.034)$ and
$\vec{A}_{11}\simeq(0.676,0.301,0.023)$  will maximize the $e-\mu$ and
$e-e$ couplings, respectively. 
In order to eventually suppress the $e-e$ coupling, we
keep the components of $\vec{A}_{12}$ that are perpendicular to
$\vec{A}_{11}$, i.e
\[
\vec{A}_{12}-\vec{A}_{11} \frac{\vec{A}_{11} \cdot \vec{A}_{12}}{|\vec{A}_{11}|^2} 
= \vec{A}_{12}+0.201 \vec{A}_{11}
\propto (-0.404,0.912,-0.064)\;.
\]
We normalize the charges by marginalizing the limit of $B(\mu \to
e \gamma)$, shown in Table~\ref{tab:cx}, then obtain 
$(Q'^{(e)}_L,Q'^{(\mu)}_L,Q'^{(\tau)}_L)=1/10\times
(-0.404,0.912,-0.064)$ and so $B(\mu \to e \gamma)=4.4\times 10^{-13}$.
Furthermore, if the right-handed charge-lepton couplings $g^l_R$ are 
nonzero, the tau-mass insertion terms in Eq.~\ref{eq7} will enhance
 $B(\mu \to e \gamma)$. We therefore simply set $g^l_R=0$.

\item [(ii)] 
Considering the quark-lepton sector strong experimental limits come
from $\mu-e$ conversion, such as $B(\mu{\rm Ti} \to e{\rm Ti})$.  
Nevertheless, these constraints can be alleviated by choosing the couplings of
$Z'q\bar{q'}$ to be
axial-vector-like from the expressions of $\mu-e$ conversion
in Sec.~\ref{mueconversion}. Therefore, we choose
${\epsilon^{u}_{L}}=-{\epsilon^{u}_{R}}$ and
${\epsilon^{d}_{L}}=-{\epsilon^{d}_{R}}$.
As the $U(1)'$ charges for the quark sector are
flavor-universal, the couplings of $Z'q\bar{q'}$ remain axial-vector-like 
under an unitary transformation of quark basis.
Recently, the LHCb Run-I data showed some deviations in $B$-meson decays
from the SM predictions:
$R_K\equiv B(B \to K\mu^+\mu^-)/B(B \to Ke^+e^-)=
0.745^{+0.090}_{-0.074}(\rm stat)\pm 0.036(\rm syst)$~\cite{bsll} has 
$2.6\sigma$ departure from unity. 
The angular observables in $B \to K^*\mu^+\mu^-$ deviate 
from the SM expectation by about $3\sigma$~\cite{bsmumu}.  
Several $Z'$ models with non-universal charged-lepton and down-type quark
couplings can explain these anomalies~\cite{zpbsmumu,zpbsll}.  
Nevertheless, these anomalies are beyond the scope of this work and
we do not attempt to explain them.

\item[(iii)]
Attempting to produce a large enough $e^{\pm}\mu^{\mp}$ cross section at 
the LHC we tune the $g^q_{L,R}$ couplings, meanwhile satisfy the dijet limits.
Fixing $g'=1$, when $Q'^{u,d}_L=-Q'^{u,d}_R\subset [0.02, 0.23]$,  we have 
$0.5 \leq \sigma(pp\to Z') \times B(Z'\to e^{\pm}\mu^{\mp})\leq 1$ fb and 
$1.5 \leq \sigma(pp\to Z') \times B(Z'\to jj)\leq 500$ fb.
If $Q'^{u,d}_L=-Q'^{u,d}_R$ were larger than 0.23, then the dijet
cross section would be too large but the $e\mu$ production cross
section would saturate around 1 fb.  On the other hand, if
$Q'^{u,d}_L=-Q'^{u,d}_R$ were less than 0.02, the $e\mu$ production
cross section would be too small.  In Table~\ref{tab:cx},
we show that $Q'^{u,d}_L=-Q'^{u,d}_R= 0.2$ can produce 
the $\sigma(pp\to Z') \times B(Z'\to e^{\pm}\mu^{\mp})\simeq 1$ fb, while
at the same time the $ee$, $\mu\mu$, $\tau\tau$, and dijet channels 
satisfy the current LHC limits at 13 TeV.

\begin{figure}[t!]
\centering
\includegraphics[height=2.2in,angle=0]{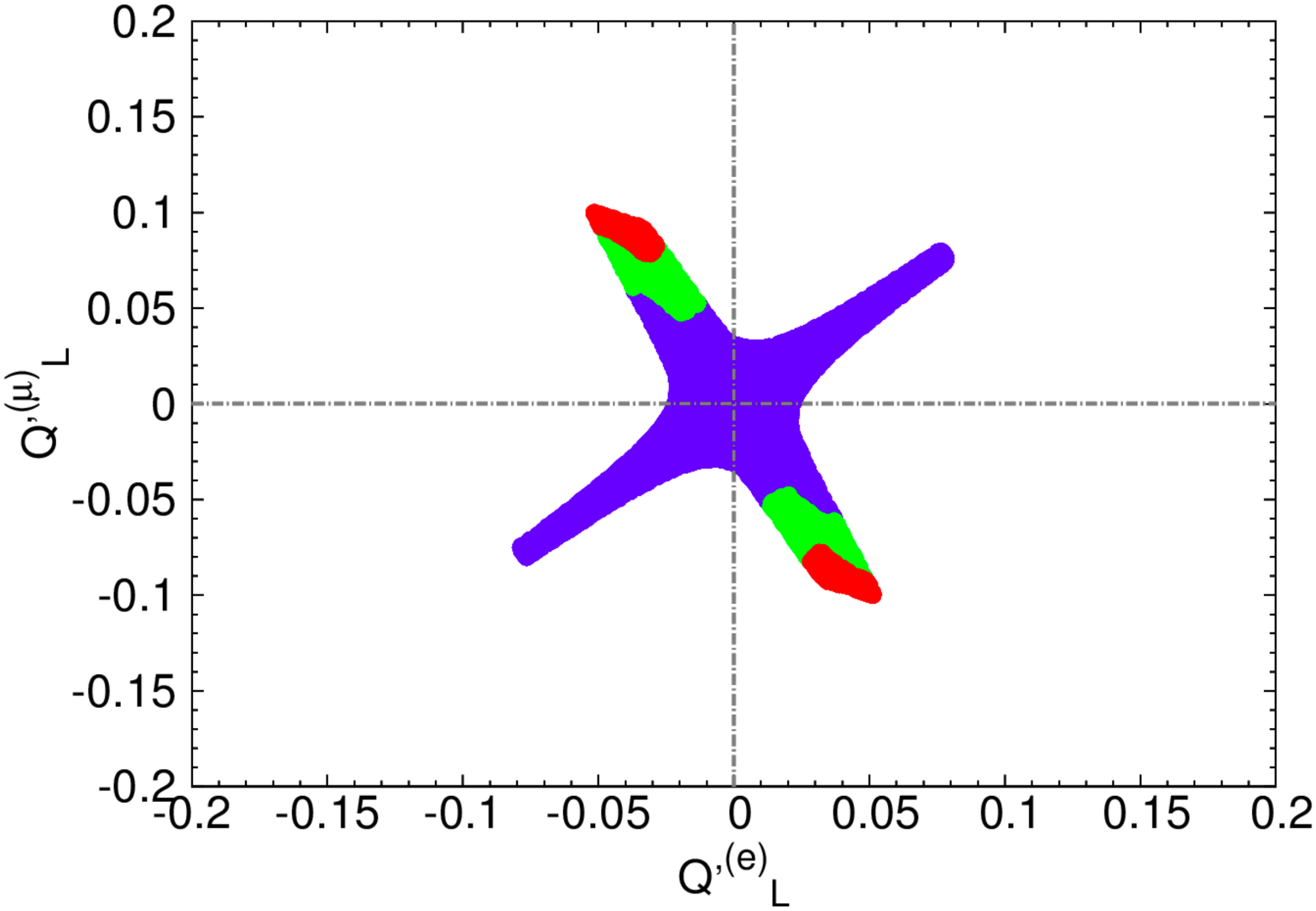}
\includegraphics[height=2.2in,angle=0]{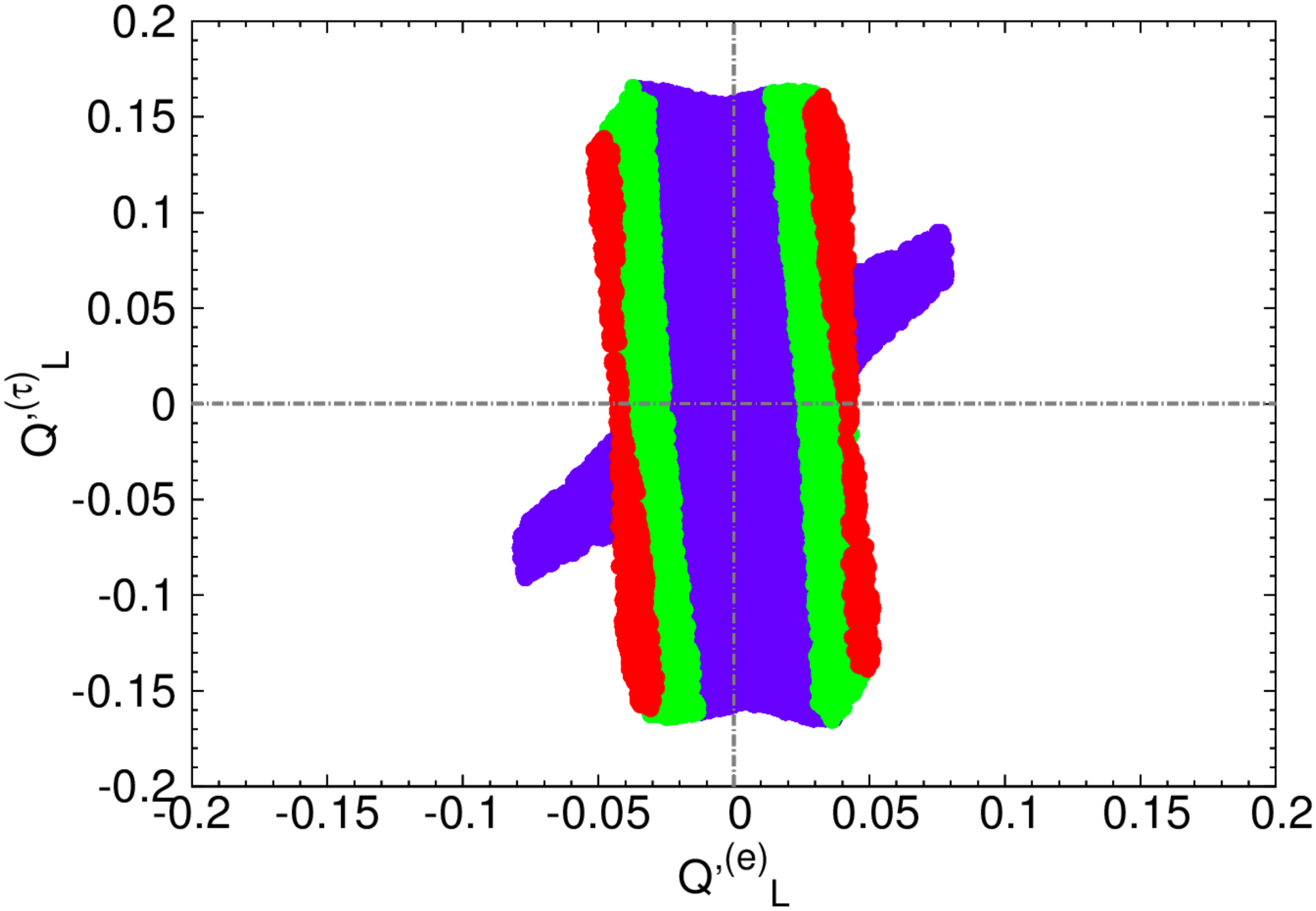}
\includegraphics[height=2.2in,angle=0]{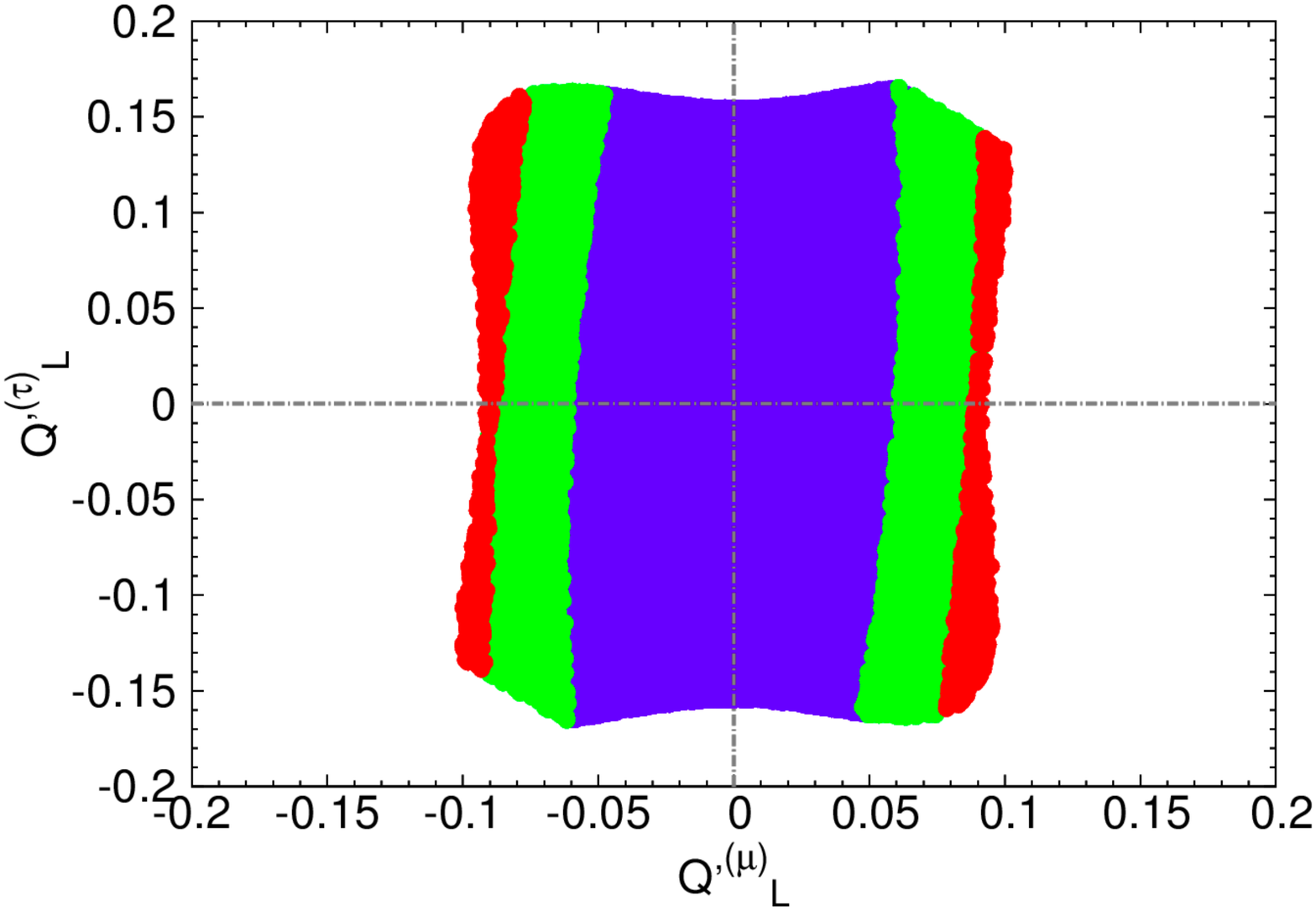}
\includegraphics[height=2.2in,angle=0]{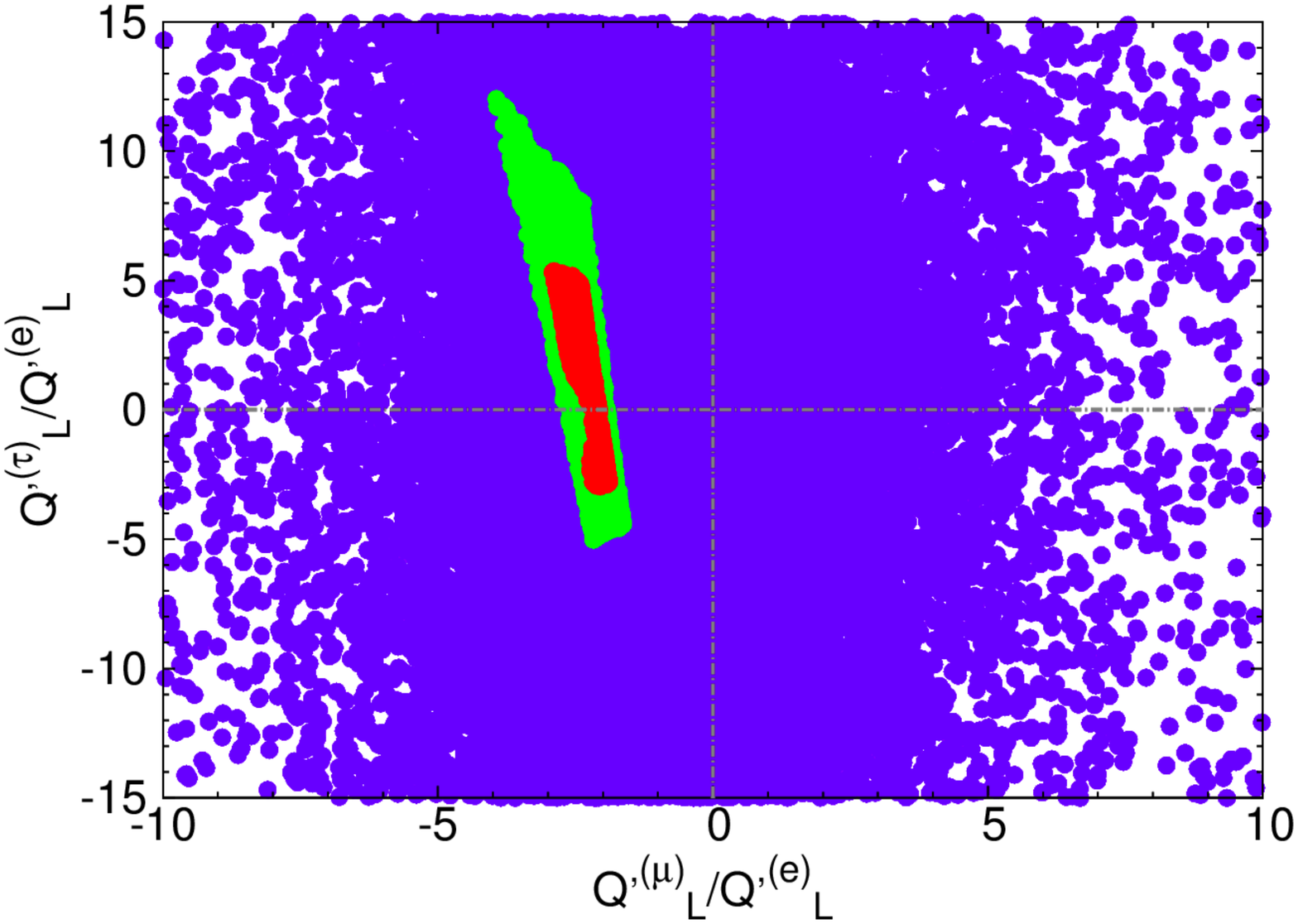}
\caption{\small \label{f1} 
Scanning over $Q'^{(e)}_L$, $Q'^{(\mu)}_L$, and $Q'^{(\tau)}_L$, while
fixing $g'=1$,
${\epsilon^{u,d}_{L}}=-{\epsilon^{u,d}_{R}}=diag(0.2,0.2,0.2)$, and
${\epsilon^{l}_{R}}=0$.  The colored points satisfy all the experimental
limits listed in Table~\ref{tab:cx}, except for $\sigma(pp\to Z')
\times B(Z'\to e^{\pm}\mu^{\mp})\simeq$ 1 $\sim$ 2 fb.  Blue points:
$\sigma(pp\to Z') \times B(Z'\to e^{\pm}\mu^{\mp})\leq$ 0.5 fb, are
the majority.  Green points: 0.5 $\leq\sigma(pp\to Z') \times B(Z'\to
e^{\pm}\mu^{\mp})\leq$ 1.0 fb.  Red points: 1.0 $\leq\sigma(pp\to Z')
\times B(Z'\to e^{\pm}\mu^{\mp})\leq$ 2.0 fb are the minority.  }
\end{figure}

\end{itemize}

Checking whether $(Q'^{(e)}_L,Q'^{(\mu)}_L,Q'^{(\tau)}_L)=1/10\times
(-0.404,0.912,-0.064)$ is the only solution or not,
we perform a scan over the parameter space of 
the three $U(1)'$ charges for the charge leptons,
$(Q'^{(e)}_L,Q'^{(\mu)}_L,Q'^{(\tau)}_L)$,  meanwhile fix $g'=1$,
${\epsilon^{u,d}_{L}}=-{\epsilon^{u,d}_{R}}=diag(0.2,0.2,0.2)$ , and
${\epsilon^{l}_{R}}=0$. The resulting scan is shown in 
in Fig.~\ref{f1}.
The $Z'$ production cross section is only
relevant to $g'$ and $Q'^{(u,d)}_{L,R}$.  From the upper-left panel in
Fig.~\ref{f1}, projecting onto
 the $(Q'^{(e)}_L,Q'^{(\mu)}_L)$ plane, we find that two
preferred directions can satisfy the experimental limits: one is along
$(\mp0.404,\pm0.912)$, which gives a large enough $\sigma(pp\to Z')
\times B(Z'\to e^{\pm}\mu^{\mp})$ at
$(Q'^{(e)}_L,Q'^{(\mu)}_L)\simeq(\mp0.404,\pm0.912)$; and the other 
one is along
$(Q'^{(e)}_L,Q'^{(\mu)}_L)\propto(1,1)$.
If we further combine with the information of the upper-right panel in 
Fig.~\ref{f1}, the latter one
corresponds to the universal $U'(1)$ charges, i.e
$(Q'^{(e)}_L,Q'^{(\mu)}_L,Q'^{(\tau)}_L)\propto(1,1,1)$, explaining
why $\sigma(pp\to Z') \times B(Z'\to e^{\pm}\mu^{\mp})$ 
would not be large along this direction.
The former one covers the benchmark point in Table~\ref{tab:cx} and
justifies the above steps in assigning the $U(1)'$ charges. There are
two solutions, $(Q'^{(e)}_L,Q'^{(\mu)}_L)\simeq(\mp0.404,\pm0.912)$,
which give a $e^{\pm}\mu^{\mp}$ production cross section larger than 1 fb, but
have weaker correlation with $Q'^{(\tau)}_L$.
The bottom-right panel of Fig.~\ref{f1} shows the ratios of
$Q'^{(\mu)}_L/Q'^{(e)}_L$ and $Q'^{(\tau)}_L/Q'^{(e)}_L$.  Requiring
the production cross section $e^{\pm}\mu^{\mp}$ larger than 1 fb
strongly confines the ratio between $U(1)'$ charges of $e$ and $\mu$,
$Q'^{(\mu)}_L/Q'^{(e)}_L\subset[-3.0,-1.8]$, but the ratio between
$U(1)'$ charges of $e$ and $\tau$ can vary a lot,
$Q'^{(\tau)}_L/Q'^{(e)}_L\subset[-3.0,5.5]$.

\section{Summary}
\label{sviii}

We have performed an analysis using a $Z'$ boson of mass 2.1 TeV 
with universal couplings to quarks but non-universal couplings to 
left-handed charged leptons in order to explain the single event 
with opposite charges and different-flavors observed by the 
ATLAS~\cite{atlas_emu}. 
The flavor-changing interactions of the $Z'$ in the charged-lepton sector 
originate from non-universal couplings in the interaction basis 
and the mass matrix is not diagonal under the flavor basis. 
In order to make this $Z'$ model more predictive, we have assumed that 
the entire lepton mixing comes from the charged-lepton sector, instead of 
the neutrino sector.  For simplicity, on one hand the $Z'$ has universal and
axial-vector-like couplings to quarks, while on the other hand, for dodging the
stringent constraints from the $\mu-e$ conversion in heavy nucleus
experiments.  Therefore, the only degrees of freedom are the gauge
coupling $g'$, three $U'(1)$ charges for charge-leptons, and one
universal $U'(1)$ charges for quarks.

We assign the $U'(1)$ charges, ${\epsilon^{l}_{L}}=1/10\times
diag(-0.404,0.912,-0.064)$, for the left-handed charged leptons to enhance
the $Z'\mu e$ but suppress $Z'ee$ couplings. The other strategies: the 
couplings to right-handed charged leptons are set to zero, and
opposite-sign charges for left- and right-handed quarks, can dodge the
observational bounds. 
We have shown a solution in Table~\ref{tab:cx} for the
normal hierarchy (NH) that a narrow-width $Z'$ boson can produce a large
enough cross section for  $\sigma(pp \to X)\times B(X \to
e^{\pm}\mu^{\mp})$ and at the same time satisfies several stringent
constraints for flavor-violating processes. Similar
solutions can be found for the inverse hierarchy(IH) case.

We have performed a scan over three $U(1)'$ charges for the charged leptons
and fixed other parameters in Fig~\ref{f1}.  It turns out that the solution in
Table~\ref{tab:cx} is quite representative. Requiring the
$e^{\pm}\mu^{\mp}$ production cross section larger than 1 fb will
restrict the ratios among $U(1)'$ charges,
$Q'^{(\mu)}_L/Q'^{(e)}_L\subset[-3.0,-1.8]$ and
$Q'^{(\tau)}_L/Q'^{(e)}_L\subset[-3.0,5.5]$.

\section*{\bf Acknowledgments}
W.-Y. K. thanks the National Center of Theoretical Sciences and 
Academia Sinica, Taiwan, R.O.C. for hospitality.
This research was supported in parts 
by the Ministry of Science and Technology (MoST) of Taiwan
under Grant Nos. 102-2112-M-007-015-MY3 and by 
US DOE under Grant No. DE-FG-02-12ER41811.


\end{document}